# Solar-Sail Deep Space Trajectory Optimization Using Successive Convex Programming

Yu Song, Shengping Gong*

School of Aerospace Engineering, Tsinghua University, Beijing, China

**Abstract:** This paper presents a novel methodology for solving the time-optimal trajectory optimization problem for interplanetary solar-sail missions using successive convex programming. Based on the non-convex problem, different convexification technologies, such as change of variables, successive linearization, trust regions and virtual control, are discussed to convert the original problem into the formulation of successive convex programming. Because of the free final-time, successive linearization is performed iteratively for the nonconvex terminal state constraints. After the convexification process, each of problems becomes a convex problem, which can be solved effectively. An augmented objective function is introduced to ensure the convergence performance and effectiveness of our algorithm. After that, algorithms are designed to solve the discrete sub-problems in a successive solution procedure. Finally, numerical results demonstrate the effectiveness and accuracy of our algorithms.

Key words: trajectory optimization, solar sailing, successive convex optimization, direct method, indirect method

## I. INTRODUCTION

In deep space, the solar radiation pressure force is one of the predominant perturbing forces exerted on the spacecraft. It can be utilized for orbit maneuvers and to achieve propellant-free interplanetary transfers. When a spacecraft utilizes solar radiation pressure as propulsion, it is called solar sailing. Early comparisons of solar sailing with chemical and ion propulsion systems showed that solar sails could match or outperform these systems for a range of mission applications, though the level of assumed technology status is of course crucial in such comparisons (MacNeal 1972). After being studied for

Corresponding author: S. Gong.
Y. Song, Ph.D. student, School of Aerospace Engineering, Tsinghua University (yumail2011@163.com).
S. Gong, Associate professor, School of Aerospace Engineering, Tsinghua University (gongsp@tsinghua.edu.cn).

several decades, a range of mission applications for solar sailing were proposed, such as interplanetary travel (Vulpetti et al. 2015), space weather forecasting (Heiligers and Mcinnes 2014), planet sample return (Hughes et al. 2006), earth and lunar observation (Heiligers et al. 2018), and even applications for non-Keplerian orbits (Gong and Li 2015a; Heiligers et al. 2016; McInnes 1999; Song et al. 2016). The main fields of the study for solar sailing are divided into attitude maneuvers and orbit and trajectory design (Dachwald 2004; Gong and Li 2015b; McInnes 2004; Mengali and Quarta 2009; Wie and Murphy 2007; Zhukov and Lebedev 1964). When considering the trajectory design, the transfer time is usually minimized because of propellant-free property of the solar sail. Numerical methods for solving the optimization problem are basically divided into two major classes as indirect methods and direct methods (Rao 2009). The indirect method requires the use of the maximum principle to derive the first order necessary conditions of the optimal solution, which leads to a two-point boundary value problem and can be solved numerically, such as shooting method(Pontryagin 1987). As summarized by Bryson et al. (Bryson and Ho 1975), the main difficulty of this method is to find a first and accurate estimate of the initial unknown costate values that produces a solution reasonably close to the boundary conditions and first-order constraints, because the extremal solutions are often very sensitive to small changes of the initial constate values. On the other hand, a direct method usually discretizes the state and control of the problem in some manner and transforms the optimal control problem into a nonlinear programming problem (NLP), which has the disadvantage of requiring a high number of parameters for the discretization process (Betts 1994; Margraves et al. 1987).

Recently, convex programming has become a popular method in the field of aerospace guidance, navigation and control (Liu et al. 2017). Because of the mature theories and numerical algorithms, as well as the high computational efficiency and polynomial complexity, convex programming has been utilized in many applications in the aerospace field such as planetary and asteroids powered landing(Acikmese and Ploen 2007; Blackmore et al. 2010; Liu 2017; Pinson and Lu 2016; Pinson and Lu 2015; Yang et al. 2017), spacecraft rendezvous and proximity operations (Liu and Lu 2013; Lu and Liu 2013), formation flight(Guo et al. 2016; Tillerson et al. 2002) and spacecraft trajectory optimization(Harris and Açıkmeşe 2014; Liu et al. 2015; Zhang 2015). Considerable research has been conducted for short-duration trajectory optimization and the effectiveness of convex programming has been fully proved in comparison with traditional methods. For the transfer trajectory optimization of low-thrust spacecraft, Tang (Tang et al. 2018) proposed a successive convex programming method and provided an accurate estimation of the initial costates of the indirect method. However, to the authors' best knowledge, there is no publication available in the literature that addresses the convex programming for the optimization of solar-sail interplanetary transfer trajectories. In this study, we explore the possibility of utilizing convex programming for the optimization of a solar sail time-optimal transfer trajectory. Compared with the fuel-optimal and minimum-landing-error problems, the current problem has some new nonconvex elements because of the free final time. For the time-optimal problem, Harris (Harris and Açıkmeşe 2014) converted the minimum-time problem into a fixed final time problems and performed an outer loop to search for the optimal time. Yang (Yang et al. 2017) solved the reduced minimum-landing-error problem and combined extrapolating and bisection methods to search for the minimum time of flight (TOF) for asteroid landing. Inspired by the free final time trajectory planning presented in the literature (Szmuk et al. 2016) that solves the fuel-optimal powered landing problem, dynamical equations of differential form will be used in this study, and the time-related variables will serve as one of the optimization parameters, which means the minimum-time problem will be solved directly without an outer loop.

The rest of the paper will be organized as follows. In Sec. II, the non-convex problem formulation is described. The dynamic model and the constraints of a solar sail spacecraft are introduced and the time-optimal interplanetary transfer trajectory problem is formulated as a time-optimal control problem. In Sec. III, we adopt several methods to convert the original non-convex problem into a second-order cone programming (SOCP) formulation. In Sec. IV, algorithms are presented to guarantee the convergence property and the optimality. At last, numerical demonstrations will prove the validity of the method and the accuracy of the algorithms.

## II. NON-CONVEX PROBLEM FORMULATION

In this section, we will present a basic model of the optimization problem. We consider the interplanetary transfer trajectory in deep space. In addition to the solar gravitation and the solar radiation pressure force, there are many perturbation forces influence the motion of the solar sail, such as the gravity of nearby celestial bodies, and the solar wind (Dachwald and Wie 2005). To analysis the mission feasibility and verify the proposed algorithms, the following simplifications as the literature are conducted in this study. The solar sail is assumed as a flat plate, and a perfectly reflecting force model is adopted. Except the solar gravitation and the solar radiation pressure force, other perturbation forces are neglected in this study. Moreover, it is assumed that the attitude of the solar sail can be changed instantaneously.

We define the reference frame of the solar sail as follows. As shown in Fig. **1**, $\hat{r}$ is the unit vector from the sun to the sail, $\hat{h}$ is the unit vector along the direction of the orbital angular momentum vector and $\hat{t}$ is the unit vector following from the right-handed reference frame with $\hat{r}$ and $\hat{h}$. The thrust of the solar sail originates from the reflection of the sunlight off the surface. The normal direction of the sail $\hat{n}$ is described by two angles, the cone angle $\alpha$ and the clock angle $\delta$. As shown in Fig. **1**, the cone angle $\alpha$ is defined as the angle between the direction of the sunlight and the normal direction of the sail $\hat{n}$. The angle parameter $\delta$ is the angle between the direction of the orbital angular momentum and the projection of $\hat{n}$ on the plane perpendicular to $\hat{r}$. Thus, the expression of $\hat{n}$ can be written as

$$\hat{n} = \cos\alpha\,\hat{r} + \sin\alpha\cos\delta\,\hat{h} + \sin\alpha\sin\delta\,\hat{t} \tag{1}$$

The domains of the angles are as follows:

$$\alpha \in [0, \frac{\pi}{2}], \delta \in [0, 2\pi] \tag{2}$$

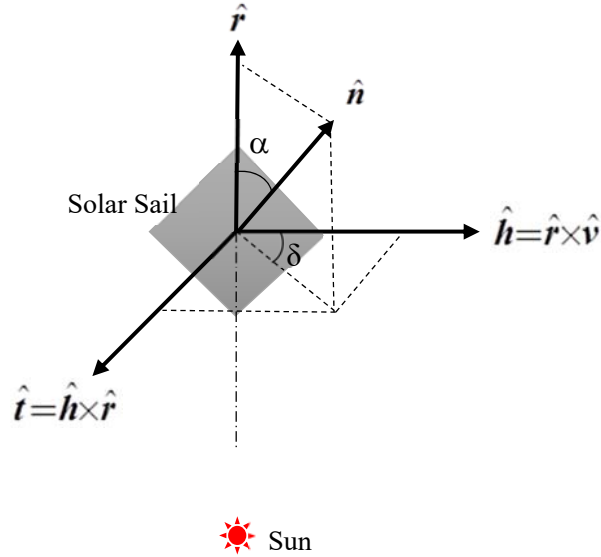

**Fig. 1 Definition of the reference frame**

The acceleration of the solar sail at a distance $r$ from the sun can be written as:

$$\boldsymbol{a}_s = \beta\frac{\mu_s}{r^2}\cos^2\alpha\,\hat{n} \tag{3}$$

where $\beta$ is the lightness number and $\mu_s$ is the solar gravitational constant (McInnes 2004).

In the heliocentric ecliptic inertial reference frame, the dynamical equations of the solar sail considering the solar radiation pressure acceleration in a two-body problem can be expressed as

$$\begin{cases} \dot{r} = v \\ \dot{v} = -\dfrac{\mu_s}{r^3} r + a_s \end{cases} \tag{4}$$

As mentioned above, the TOF of the transfer trajectory is usually used as the performance index of the optimal control problem because no fuel is consumed by the solar sail. Consider the interplanetary rendezvous problem of a solar sail. The solar sail is assumed to leave the Earth with zero hyperbolic excess energy ($C_3 \equiv 0$ km$^2$/s$^2$) and rendezvous with the target object (Mengali and Quarta 2009). With the departure time at the starting planet being $t_0$ and the rendezvous time at the target object being $t_f$, the initial and terminal state constraints can be described as:

$$\Psi(t_0) = \begin{bmatrix} r(t_0) - r_0 \\ v(t_0) - v_0 \end{bmatrix} = \mathbf{0} \tag{5}$$

$$\Psi(t_f) = \begin{bmatrix} r(t_f) - r_f \\ v(t_f) - v_f \end{bmatrix} = \mathbf{0} \tag{6}$$

where $\begin{bmatrix} r(t_0) \\ v(t_0) \end{bmatrix}$ and $\begin{bmatrix} r(t_f) \\ v(t_f) \end{bmatrix}$ are the initial and final states of the solar sail, $\begin{bmatrix} r_0 \\ v_0 \end{bmatrix}$ and $\begin{bmatrix} r_f \\ v_f \end{bmatrix}$ are the states of the departure and target celestial bodies, respectively.

Given the dynamical equations as in Eq.(1)-(4) and the constraints as in Eq.(5)-(6), the time-optimal control problem can be formulated as:

**Problem P0:**

$$\begin{aligned} \min \ & J = t_f - t_0 \\ s.t. \ & \text{Eq. (1-6)} \end{aligned} \tag{7}$$

As shown in Problem P0, the objective function is non-convex and the dynamical equations are highly non-convex. Therefore, some convexification techniques are needed to handle the problem.

### III. CONVEXIFICATION OF NON-CONVEX OPTIMIZATION PROBLEM

The optimization problem as in Eq.(7) does no have the formulation of convex optimization, which means that considerable efforts are required to convert and discretize the problem into the framwork of SOCP and then a single or a sequence of convex optimization problems can be solved to approach the solution of the original optimization problem(Liu et al. 2017). First, the change of control variables is adopted to reduce the coupling between the magnitude and direction of the solar radiation pressure acceleration. Then, we formulate the differential dynamical equations instead of the continuous-time ones. Usually, when the final time is free and if there is no appropriate alternative independent variable for continuous-time dynamical equations, an outer loop to search the optimal final time is necessary. The differential dynamical equations have an obvious advantage in solving this problem, in which the time interval can serve as an optimization parameter considering a fixed discrete number $N$. Additionally, we will apply successive convexification methods as introduced in (Liu and Lu 2014; Mao et al. 2017; Szmuk et al. 2016) to handle the non-convex elements of the problem in the nonlinear dynamics and constraints. At last, the virtual control method is adopted to overcome the artificial infeasibility.

**A. CHANGE OF VARIABLES**

As the only force for orbital maneuver, the acceleration of the solar sail shown as Eq.(3) consists of two parts: the magnitude and direction of the solar radiation pressure force. The cone angle $\alpha$ appears in both parts, which leads to difficulty of the iterative convergence. Thus, we change the variables by equivalent transformation and decouple the magnitude and direction of the acceleration as follows.

Define the new control variables $\boldsymbol{u} \triangleq [u_1 \quad u_2 \quad u_3]^T$:

$$\begin{aligned} u_1 &= \cos^3 \alpha \\ u_2 &= \cos^2 \alpha \sin \alpha \cos \delta \\ u_3 &= \cos^2 \alpha \sin \alpha \sin \delta \end{aligned} \tag{8}$$

Then we can rewrite Eq.(1) and (3) as the following form:

$$\boldsymbol{a}_s = \beta \frac{\mu_s}{r^2} \left( u_1 \hat{\boldsymbol{r}} + u_2 \hat{\boldsymbol{h}} + u_3 \hat{\boldsymbol{t}} \right) \tag{9}$$

Moreover, there exists an essential constraint $\varPhi$ as follows, which is nonlinear and will be addressed in a later discussion.

$$\varPhi(\boldsymbol{u}) = u_1^2 + u_2^2 + u_3^2 - u_1^{4/3} = 0 \tag{10}$$

**B. DISCRETIZATION**

For a specified number of time nodes $N$, the continuous-time dynamical equations can be approximated to a discrete differential form. For a given initial time $t_0$ and final time $t_f$, the continuous time domain can be replaced by a discrete time sequence from $t_0$ to $t_f$. We adopt the transformation relationship introduced in (Szmuk et al. 2016) as follows.

$$\Delta t = \frac{t_f - t_0}{N - 1} \tag{11}$$

$$k_f \triangleq N - 1$$

$$t = k\Delta t + t_0 \qquad k \in [0, k_f] \tag{12}$$

At the same time, the time interval $\Delta t$ will be an optimization parameter instead of the final time $t_f$ considering a fixed initial time $t_0$ and discrete number $N$.

For the discrete time sequence as shown above, the continuous-time dynamical equations as Eq.(4) can be reformulated as a differential form at the discrete points. The dynamics are discretized assuming that control parameters of solar radiation pressure acceleration are interpolated linearly between two adjacent time nodes. The discrete form of Problem P0 is shown as Problem P1 below, and the first two equations in Eq.(14) are in the same form as those in the literature (Szmuk et al. 2016).

**Problem P1:**

$$\min \ \Delta t \tag{13}$$

Subject to:

$$
\begin{aligned}
\boldsymbol{r}[k+1] &= \boldsymbol{r}[k] + \boldsymbol{v}[k]\Delta t + \frac{1}{3}\left(\boldsymbol{a}[k] + \frac{1}{2}\boldsymbol{a}[k+1]\right)\Delta t^2 & k \in [0, k_f] \\
\boldsymbol{v}[k+1] &= \boldsymbol{v}[k] + \frac{1}{2}\left(\boldsymbol{a}[k] + \boldsymbol{a}[k+1]\right)\Delta t & k \in [0, k_f] \\
\boldsymbol{a}[k] &= -\frac{\mu_s}{r^2[k]}\hat{\boldsymbol{r}}[k] + \boldsymbol{a}_s[k] & k \in [0, k_f]
\end{aligned}
\tag{14}$$

$$\begin{bmatrix} \boldsymbol{r}[0] - \boldsymbol{r}_0 \\ \boldsymbol{v}[0] - \boldsymbol{v}_0 \end{bmatrix} = \boldsymbol{0} \tag{15}$$

$$\begin{bmatrix} \boldsymbol{r}[k_f] - \boldsymbol{r}_f(t_f) \\ \boldsymbol{v}[k_f] - \boldsymbol{v}_f(t_f) \end{bmatrix} = \boldsymbol{0} \tag{16}$$

*Remark*1: $\begin{bmatrix} \boldsymbol{r}_f(t_f) \\ \boldsymbol{v}_f(t_f) \end{bmatrix}$ in Eq.(16) means that the terminal state constraints are changed when the final time $t_f$ is changed in the iterative process. The non-convexity between the terminal state and the optimization parameter, namely the state of the target object and the terminal time $t_f$ will be addressed in a later section.

*Remark*2: If there is no special explanation, this paper will use $k$ to represent the time nodes and $i$ to represent the number of iterations.

## C. SUCCESSIVE CONVEXIFICATION

In this section we will present a successive convexification procedure to convert the problem discussed above to the SOCP formulation. We perform the linearization to the differential dynamics shown in Eq.(10) and Eq.(14). Trust regions are used to bound the variables involved in the linearization with the previous iteration. Unlike the state constraints in the literature (Szmuk et al. 2016), the terminal state constraints of our problem are time-varying and we will adopt the forward integration method to approximate the terminal state during the iteration. Also, the virtual control mentioned in the literature will be adopted to handle the artificial infeasibility in the previous several iterations.

### C.1 SUCCESSIVE LINEARIZATION

There are two sources of nonlinearity in Problem P1 that can be linearized, the nonlinear dynamics Eq.(14) and nonlinear control constraints Eq.(10). Here we define a new variable form as introduced in (Szmuk et al. 2016):

$$\boldsymbol{\Gamma}_r[k] = \begin{bmatrix} \boldsymbol{r}^T[k] & \boldsymbol{r}^T[k+1] \end{bmatrix}^T \qquad k \in [0, k_f] \tag{17}$$

where the subscript of the variable $\boldsymbol{\Gamma}$ represents the type of the variable, for example $\boldsymbol{\Gamma}_r$ represents the position vectors at the $k^{th}$ and $k+1^{th}$ nodes.

Based on the previous definition, all the variables shown up in this problem can be written as the similar form:

$$\boldsymbol{\Gamma} = \begin{bmatrix} \Delta t & \boldsymbol{\Gamma}_r^T[k] & \boldsymbol{\Gamma}_v^T[k] & \boldsymbol{\Gamma}_u^T[k] \end{bmatrix}^T \tag{18}$$

By adopting the successive linearization method, the dynamics shown in Eq.(14) can be reformulated as follows:

$$\begin{aligned} \boldsymbol{r}[k+1] &= \boldsymbol{r}[k] + \boldsymbol{f}_r(\boldsymbol{\Gamma}_{i-1}[k]) + \boldsymbol{A}_{i-1} \cdot \delta\boldsymbol{\Gamma}[k] & k \in [0, k_f] \\ \boldsymbol{v}[k+1] &= \boldsymbol{v}[k] + \boldsymbol{f}_v(\boldsymbol{\Gamma}_{i-1}[k]) + \boldsymbol{B}_{i-1} \cdot \delta\boldsymbol{\Gamma}[k] & k \in [0, k_f] \end{aligned} \tag{19}$$

where $\boldsymbol{f}_r$ and $\boldsymbol{f}_v$ are parts of the dynamics containing the nonlinearity of variables, $\boldsymbol{A}_{i-1}$ and $\boldsymbol{B}_{i-1}$ are the Jacobian matrix based on the solutions of the $i$-$1^{th}$ iteration.

$$\begin{aligned} \boldsymbol{f}_r(\boldsymbol{\Gamma}[k]) &= \boldsymbol{v}[k]\Delta t + \frac{1}{3}\left(\boldsymbol{a}[k] + \frac{1}{2}\boldsymbol{a}[k+1]\right)\Delta t^2 \\ \boldsymbol{f}_v(\boldsymbol{\Gamma}[k]) &= \frac{1}{2}(\boldsymbol{a}[k] + \boldsymbol{a}[k+1])\Delta t \end{aligned} \tag{20}$$

$$A_{i-1} = \left.\frac{\partial f_r}{\partial \Gamma}\right|_{\Gamma_{i-1}[k]}$$
$$B_{i-1} = \left.\frac{\partial f_v}{\partial \Gamma}\right|_{\Gamma_{i-1}[k]} \quad (21)$$

The nonlinearity in control constraints in Eq.(10) will also be successively satisfied by the following linearization.

$$\bar{\Phi}(\boldsymbol{u}_{i-1}, \boldsymbol{u}) = \Phi(\boldsymbol{u}_{i-1}[k]) + \boldsymbol{C}_{i-1}\delta\boldsymbol{u}[k] = 0 \quad (22)$$

where $C_{i-1}$ is the Jacobian matrix of Eq.(10) based on the solutions $\boldsymbol{u}$ of the previous iteration as well.

$$C_{i-1} = \left.\frac{\partial \Phi}{\partial \boldsymbol{u}}\right|_{\boldsymbol{u}_{i-1}[k]} \quad (23)$$

### C.2 TRUST REGION

The potential risk during linearization is rendering the problem unbounded, and new states may deviate significantly from the nominal state and control sequence (Mao et al. 2016; Szmuk et al. 2016), which increases the difficulty of convergence severely and possibly leads to infeasibility of the linearized problem. To avoid this risk, the trust regions are adopted to constrain the variables as follows:

$$\|\delta\boldsymbol{u}\| \leq \eta_u \quad (24)$$

$$\delta\Delta t^2 \leq \eta_{\Delta t} \quad (25)$$

where $\eta_u$ and $\eta_{\Delta t}$ are trust region radius that restricts the control inputs and time interval in every iteration and will be penalized via additional terms in the augmented objective function. In the first few iterations, the trust region radius can be appropriately larger to find a proper convergence direction. With the increase of iterations, the trust region radius should be gradually narrowed considering the convergence and validity of the linearized problem.

Remarkably, as long as the control inputs $\boldsymbol{u}$ and time interval $\Delta t$ are constrained, the new states, namely the position and velocity vectors can be constrained as well due to the linearized dynamic equations.

### C.3 TERMINAL STATE CONSTRAINTS

In the previous literature, most of the terminal state constraints are time independent or in simple mapping relationship. For our problem, the constraints of terminal state as Eq.(16) are time-varying because of the integration of time. These highly nonlinear constraints can not be applied directly in a convex optimization procedure when regarding the final time as an optimization parameter. Existing literature mainly considers the final time fixed in an iteration and update the final time at the end of every iteration (Harris and Açıkmeşe 2014; Yang et al. 2017).

Considering a small variation range of the terminal time, which is reasonable under the constraint of trust region of time, the state of the target object varies slightly near the previous state. Therefore, we can use the forward integration method to approximate the new state changed because of the small time variation $\delta t_f$.

We define the state of the target object as $x_{ter} = \begin{bmatrix} r_{ter} \\ v_{ter} \end{bmatrix}$, where $r_{ter}$ and $v_{ter}$ indicate the position and velocity of the target object. The state of the target object can be approximated as follows.

$$\dot{x}_{ter} = g(x_{ter}) = \begin{bmatrix} v_{ter} \\ -\dfrac{\mu}{r_{ter}^2} r_{ter} \end{bmatrix}$$

$$x_{ter}(t_{f,i}) = x_{ter,i-1}(t_{f,i-1}) + g(x_{ter,i-1}(t_{f,i-1}))\delta t_{f,i}$$

(26)

where the nonlinear function $g$ represents the two-body motion equation of the target object, $x_{ter,i-1}(t_{f,i-1})$ is the nominal terminal state calculated accurately using the obtained terminal time $t_{f,i-1}$ from the previous iteration.

*Remark*: To avoid the error accumulation caused by the approximation, the nominal terminal state needs to be updated every iteration as long as the terminal time $t_{f,i}$ is obtained in the $i$th iteration.

### C.4 AUGMENTED OBJECTIVE FUNCTION

Considering the linearization and trust regions adopted to our problem, the objective function Eq.(13) in Problem P1 is apparently insufficient. To ensure the convergence performance and the effectiveness of our algorithm, take into account the augmented objective function as follows:

$$\min \ \Delta t + w_{\Delta t} \cdot \eta_{\Delta t} + w_u \cdot \eta_u \tag{27}$$

where $\eta_u$ and $\eta_{\Delta t}$ are trust region radius mentioned previously and $w_{\Delta t}$ and $w_u$ are the penalty parameters respectively, which will be updated by a classical L1 penalty method (Nocedal and Wright 2006).

### D. ARTIFICIAL INFEASIBILITY

Because of the linearization and state constraints mentioned previously, infeasibility occurs in the first few iterations and obstructs the iterative process, namely artificial infeasibility (Mao et al. 2016). There are several methods of dealing with the problem. One is to relax some of the constraints, such as terminal constraints, and penalizing the relaxing radius in the augmented objective function (Liu et al. 2015). For our problem considering a free terminal time, relaxed terminal constraints mean more algorithm complexity. Another commonly used method is virtual control (Mao et al. 2016; Szmuk et al. 2016). We introduce an additional acceleration term $a_v$ into the dynamical equations as the virtual control input:

$$a = -\frac{\mu_s}{r^2}\hat{r} + a_s + a_v \tag{28}$$

Then we further augment the objective function as follows:

$$\min \quad \Delta t + w_{\Delta t} \cdot \eta_{\Delta t} + w_u \cdot \eta_u + w_{a_v} \cdot \|a_v\| \tag{29}$$

where $w_{a_v}$ is a heavy penalty factor for the magnitude of virtual control input.

The virtual control input $a_v$ is significant at the first few iterations to prevent the artificial infeasibility and will approach zero after several iterations because of the heavy penalty in the objective function. Namely, the virtual control input will not make the solution deviate from the original problem while increasing the problem's convergence performance.

After the successive convexification processes, we obtain the following convex optimization problem configurations as Problem P2. The algorithms and numerical demonstration discussed in the following sections are based on the formulation of Problem 2.

**Problem P2:**

Objective Function:

$$\min \quad \Delta t + w_{\Delta t} \cdot \eta_{\Delta t} + w_u \cdot \eta_u + w_{a_v} \cdot \|a_v\| \tag{30}$$

Boundary Conditions:

$$x[0] = \begin{bmatrix} r_0 \\ v_0 \end{bmatrix} \tag{31}$$

$$x[k_f] = x_{\text{ter},i-1}[k_f] + g(x_{\text{ter},i-1}[k_f])\delta\Delta t_i \cdot (N-1) \tag{32}$$

Dynamics:

$$\boldsymbol{\Gamma} = \begin{bmatrix} \Delta t & \boldsymbol{\Gamma}_r^T[k] & \boldsymbol{\Gamma}_v^T[k] & \boldsymbol{\Gamma}_u^T[k] \end{bmatrix}^T \tag{33}$$

$$\begin{aligned} r[k+1] &= r[k] + f_r(\boldsymbol{\Gamma}_{i-1}[k]) + A_{i-1} \cdot \delta\boldsymbol{\Gamma}[k] \quad k \in [0, k_f] \\ v[k+1] &= v[k] + f_v(\boldsymbol{\Gamma}_{i-1}[k]) + B_{i-1} \cdot \delta\boldsymbol{\Gamma}[k] \quad k \in [0, k_f] \end{aligned} \tag{34}$$

$$a[k] = -\frac{\mu_s}{r^2[k]}\hat{r}[k] + a_s[k] + a_v[k] \qquad k \in [0, k_f] \qquad (35)$$

$$\begin{aligned} f_r(\Gamma[k]) &= v[k]\Delta t_{i-1} + \frac{1}{4}(a[k] + a[k+1])\Delta t_{i-1}^2 \\ f_v(\Gamma[k]) &= \frac{1}{2}(a[k] + a[k+1])\Delta t_{i-1} \end{aligned} \qquad (36)$$

Control Constraints:

$$\begin{aligned} 0 &\leq u_1 \leq 1 \\ -1 &\leq u_2 \leq 1 \\ -1 &\leq u_3 \leq 1 \end{aligned} \qquad (37)$$

$$\overline{\Phi}(u_{i-1}, u) = 0 \qquad (38)$$

## IV. SUCCESSIVE SOLUTION PROCEDURE

In this section, we will propose the successive solution procedure to find the solution of the optimal control problem by solving the convex problem iteratively. For the first iteration, we need provide an initial reference solution for the iterative process. Many works in the literature have shown that the successive solution method is insensitive to the initial reference solution(Mao et al. 2017; Szmuk et al. 2016). We will illustrate this performance in the next section. In our successive solution procedure, the $i^{th}$ solution is used to approximately linearize the nonlinear dynamics and constraints in the $i+1^{th}$ iteration. To guarantee the convergence of the iterative process, a kind of quadratic penalty method is adopted to update the penalty parameters. We perform the process repeatedly until the convergence criteria is satisfied. The procedure is stated in Algorithm 1.

| **Algorithm 1** | Successive Solution Algorithm |
|---|---|

**Step 1**: Select an initial reference solution $\{x_0; u_0\}^*$. Specify the initial guess of the terminal time $t_{f,0}$ and compute $\Delta t_0$ using the discrete number $N$;

**Step 2**: Initialize the weight coefficients $\{w_{\Delta t}, w_u, w_{a_v}\}$.

**Step 3:** Perform the first iteration and get the solution $x_1$, $u_1$ and $\Delta t_1$. Update the weight coefficients $\{w_{\Delta t}, w_u, w_{a_v}\}$.

**Step 4**: For the $i^{th}$ iteration, update the reference result with the $i$-$1^{th}$ iteration solution $x_{i-1}$, $u_{i-1}$ and $\Delta t_{i-1}$ and solve Problem 2;

**Step 5**:

    **If** $\|u_i - u_{i-1}\| \leq \varepsilon_u$ and $|\Delta t_i - \Delta t_{i-1}| \leq \varepsilon_{\Delta t}$,

    return $\{\boldsymbol{x}_i; \boldsymbol{u}_i; \Delta t_i\}$ as the result of Problem P2;

  **else**

    Update $\{w_{\Delta t}, w_u, w_{a_v}\}$;

    $i=i+1$;

    and back to **Step 4**;

  **end**

**Step 6**: If $i \geq i_{max}$, return "Infeasible" for the problem.

*The subscript of $\boldsymbol{x}_0$ and $\boldsymbol{u}_0$ indicate the initial reference solution for the first iteration.

## V. NUMERICAL TEST CASES

In this section, we will introduce the numerical results and demonstrate the effectiveness and robustness of the proposed algorithm for solar sail transfer trajectory optimization. To demonstrate the effectiveness of our method, the existing methods such as the indirect method and the direct pseudospectral method will be used to solve the same cases for comparison. Guaranteed by the first order necessary conditions, solutions of indirect methods are considered as the local optimal solutions. Repeatedly solved solutions of the indirect method are considered as the global optimal solutions (He et al. 2014). In addition, the direct pseudospectral method has been widely used for solar sailing trajectory optimization in the literature (Heiligers et al. 2015; Melton 2002). For the verification of the effectiveness, accuracy and optimality of the proposed algorithm, solutions obtained by the indirect method and direct pseudospectral method are referenced and compared with the solution of convex programming.

To verify the robustness of the proposed algorithm, different initial guesses of the terminal time will be set and the results will be compared to illustrate the convergence properties of the algorithm.

Moreover, test cases that have been studied abundantly in the literature, such as the rendezvous problem with Mars and asteroid Apophis (Hughes et al. 2006; Mengali and Quarta 2009), will be considered to verify the applicability and effectiveness of the proposed method.

### A. SELECTION OF CELESTIAL BODIES AND PARAMETERS

We consider the orbit transfer trajectories from Earth with zero hyperbolic excess energy and the time to leave the Earth is fixed. The characteristic acceleration of the solar sail is set to $a_c = 0.5$ mm/s$^2$ (corresponding to a lightness number of $\beta$=0.0843), which is reasonable and widely used in the literature (Hughes et al. 2006; Mengali and Quarta 2009).

For the sake of computational efficiency, all distance quantities are normalized with the astronomical unit (AU) and time-related quantities are normalized with $1/2\pi$ year. The reference orbit elements from JPL's database[†] at MJD 57800 are listed in Table **1**.

**Table 1 Reference orbit elements of asteroids**

|  | $a$ (AU) | $e$ | $i$ (rad) | $\Omega$ (rad) | $\omega$ (rad) | $M$ (rad) |
| --- | --- | --- | --- | --- | --- | --- |
| Earth | 0.9995 | 0.0166 | 0.0000 | 3.5798 | 4.4677 | 0.7822 |
| Venus | 0.7233 | 0.0067 | 0.0592 | 1.3375 | 0.9602 | 6.1517 |
| Mars | 1.5237 | 0.0935 | 0.0322 | 0.8640 | 5.0032 | 1.0011 |
| Apophis[*] | 0.9222 | 0.1911 | 0.0581 | 3.5684 | 2.2060 | 3.7619 |

[*] The reference orbital elements of asteroid Apophis are defined at Modified Julian Date (MJD) 54441

To use the successive solution method for convex programming, the initial reference values of the iterative process need to be specified. As described in the previous section, the initial reference solution can be selected roughly and will not affect the convergence performance. Hence, the initial reference state variables are chosen as the state variables of the initial celestial body and the control angles are specified as constants, for example $\alpha \equiv 30°$ and $\delta \equiv 180°$. Another initial reference variable needs to be specified is the terminal time. Once the time of departure time is determined, the guess of the final time will be estimated conservatively according to the difference of the orbital elements and the performance of the solar sail.

For the indirect method, the original optimal control problem will be converted to a two-point boundary value problem and will be solved by shooting method. The initial guess of the unknown values are randomly generated until the final convergence. For the direct pseudospectral method, the initial guesses will be set to the same as those of the convex optimization.

The MATLAB software CVX, a package for specifying and solving convex programs (Grant and Boyd 2015), will be used and the solver ECOS (Domahidi et al. 2013) is called in our numerical test cases. When performing the indirect shooting procedure, MinPack-1 (Moré et al. 1980), a package of Fortran subprograms for the numerical solution of systems of nonlinear equations and nonlinear least squares problems is used. In addition, the general-purpose software GPOPS (Rao et al. 2010) for solving optimal

---

[†] Data available online at https://ssd.jpl.nasa.gov/dat/ELEMENTS.NUMBR [retrieved 22 February 2017].

control problems will be used to obtain results of the direct pseudospectral method.

All the numerical computations are executed on a PC with an Intel Core i7-4470 at 3.4GHz and 8GB of memory.

## B. NUMERICAL RESULTS

### B.1 RENDEZVOUS WITH MARS

As first test case, we consider the transfer mission from Earth to Mars. The departure time is fixed to MJD 55840 (Oct. 06, 2011), which is the same as the case in the literature (Hughes et al. 2006). In this case, a value of $N=100$ is used for the discretization described in Sec. III.A. The value of other parameters used for this case are listed in Table **2**.

**Table 2 Numerical Simulation Parameters**

| Parameter | Value | Units |
|---|---|---|
| $t_0$ | 55840 | MJD |
| $t_{f,0}$ | 56140 | MJD |
| $\beta$ | 0.0843 | - |
| $N$ | 100 | - |
| $\eta_{u,\max}$ | 0.5 | - |
| $\eta_{\Delta t,\max}$ | 10 | day |
| $w_{\Delta t}$ | 0.1 | s$^{-1}$ |
| $w_u$ | 0.01 | - |
| $w_{av}$ | $1\times 10^7$ | - |
| $\varepsilon_u$ | $1\times 10^{-5}$ | - |
| $\varepsilon_{\Delta t}$ | $1\times 10^{-3}$ | day |

For the given departure time at Earth, the solar sail rendezvouses with Mars on May 5 2013, and the TOF is 577.591 days. The result is consistent with the results in the literature (Hughes et al. 2006), which verifies the optimality of the proposed method. The indirect method is used to solve the problem with then same boundary conditions for comparison. The results obtained by successive convex programming and indirect method is as Table 3, Fig. 2 and Fig. 3.

**Table 3 Convergence Solution of the Mars Rendezvous Problem**

| Description | Result |
|:---:|:---:|
| Number of iterations | 16 |
| Departure date | 10-06-2011 |
| TOF(day) | 577.391 |
| Rendezvous date | 05-05-2013 |
| Iteration error of control | $3.61\times10^{-5}$ |
| Rendezvous error(AU) | $1.22\times10^{-9}$ |

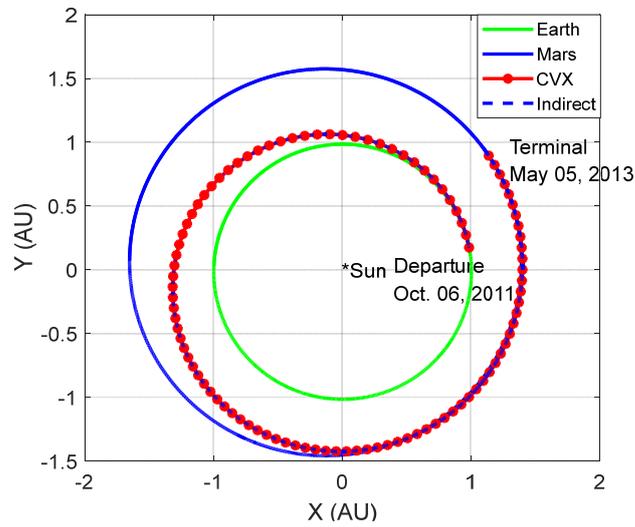

Fig. 2 Transfer trajectory of Mars rendezvous

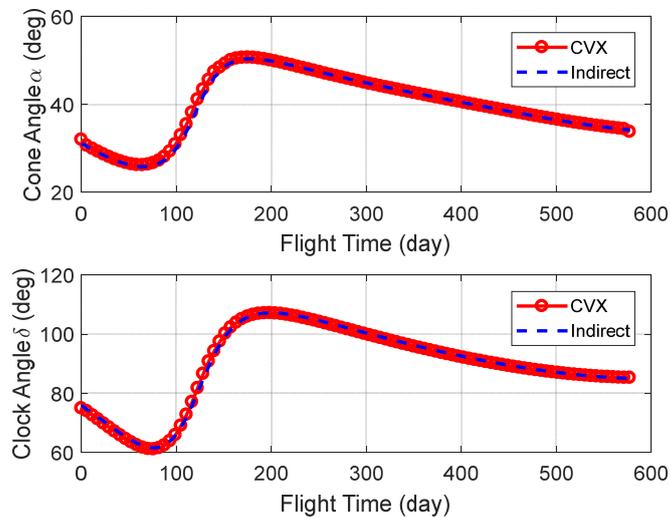

**Fig. 3 Control profiles of Mar rendezvous**

The parameters listed in Table **3** represent the results of the solution obtained using convex programming. The initial guess for the TOF is 300 days. The convergence criterion we used is the iteration error that represents the error of the control variables between two adjacent iterations. The algorithm converges after 16 times iteration. After we obtain the solution, the reintegration procedure is conducted to verify the solution. The difference in the state variables between the solar sail and Mars at the time of rendezvous after reintegration is of $1.22\times10^{-9}$ normalized time unit, which indicates that the obtained solution by convex optimization is valid and accurate enough.

Compared to the solution obtained with the indirect method, the TOF obtained by convex programming is slightly longer because of the discretization error. The position and velocity of the solar sail in the transfer trajectory is basically in agreement with the result of the indirect method, as shown in Fig. **2**. The control angles of the solar sail are also consistent with the results of the indirect method shown as Fig. **3**.

The numerical results above confirm the effectiveness and optimality of the proposed algorithm for convex programming. In the next section the convergence and robustness performance will be discussed by changing the simulation parameters in different conditions.

**B.2 RENDEZVOUS WITH ASTEROID ASTEROIDS**

To demonstrate the efficiency of the proposed method, the direct pseudospectral method will be adopted for comparison. In this case, the asteroid 99942 Apophis is chosen as the target object. The time of departure is set as MJD 56258 (Nov. 27, 2012) according to the test case in another literature (Mengali and Quarta 2009). For the given time of departure time, the solutions obtained by convex programming and direct pseudospectral method are as Table 4, Fig. 4 and Fig. 5.

**Table 4 Convergence Solution of the Apophis Rendezvous Problem**

| Description | Result | |
| --- | --- | --- |
| | CVX | GPOPS |
| Number of iterations | 16 | 20 |
| TOF(day) | 279.912 | 279.945 |
| Effective computation time (second) | 0.960 | 28.240 |

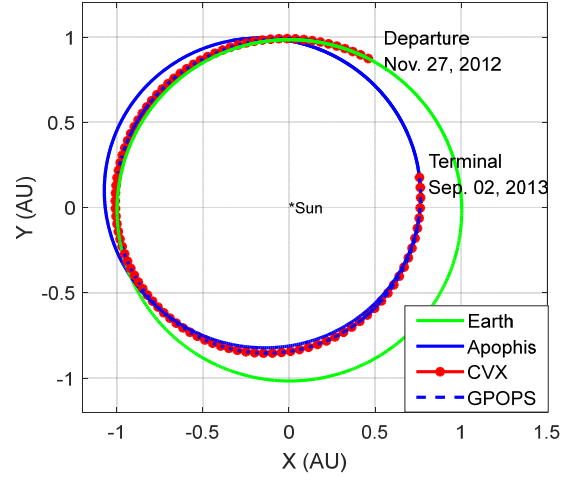

Fig. 4 Transfer trajectory of Apophis rendezvous

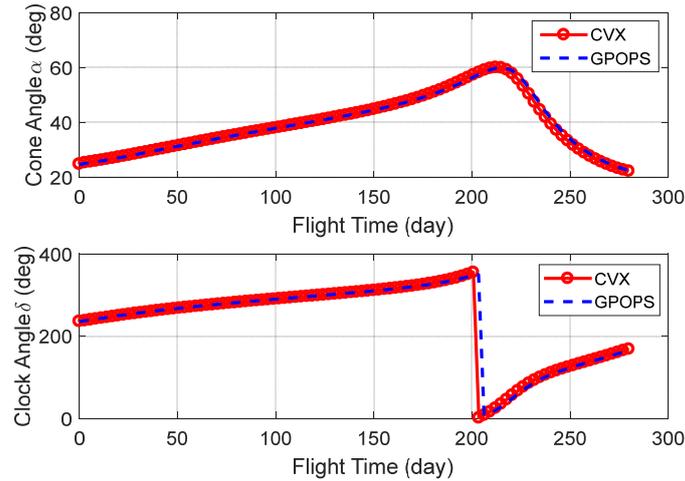

Fig. 5 Control profiles of Apophis rendezvous

The initial reference and boundary conditions are set to the same situation of the two methods. The results and computation time of each method are recorded. As shown in **Table 4**, the results for the TOF obtained by convex optimization and direct pseudospectral method are almost identical except for minor error. The trajectory in **Fig. 4** and the control profiles in **Fig. 5** indicate that the results of the different method are coincident. Compared with the direct method, the proposed method are more efficient.

**B.3 RENDEZVOUS WITH VENUS**

In this case, we select the Venus as the target object. To verify the convergence and robustness performance of the proposed algorithm, different conditions are set for further discussion in this case.

The solution obtained by convex programming is as follows. The solar sail departure from the Earth on October 26 2019 and rendezvouses with Venus on August 02 2020, corresponding to a TOF of 281.167 days. The transfer trajectory and control profiles are shown in **Fig. 6** and **Fig. 7**.

**Table 5 Convergence Solution of the Venus Rendezvous Problem**

| Description | Result |
|---|---|
| Number of iterations | 12 |
| Departure time | 10-26-2019 |
| TOF(day) | 281.167 |
| Rendezvous Time | 08-02-2020 |

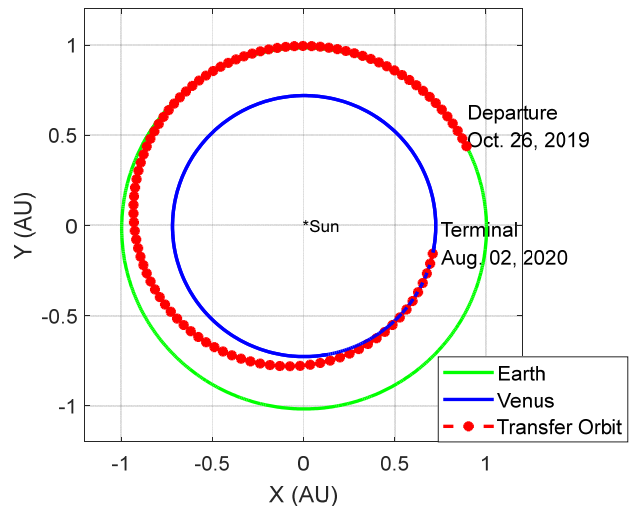

**Fig. 6 Transfer trajectory of Venus rendezvous**

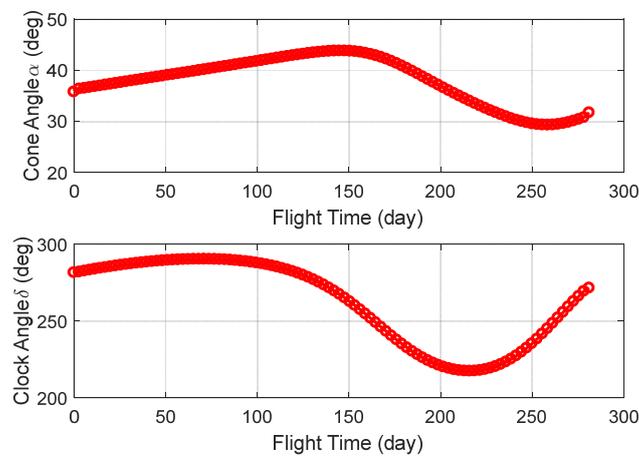

**Fig. 7 Control profiles of Venus rendezvous**

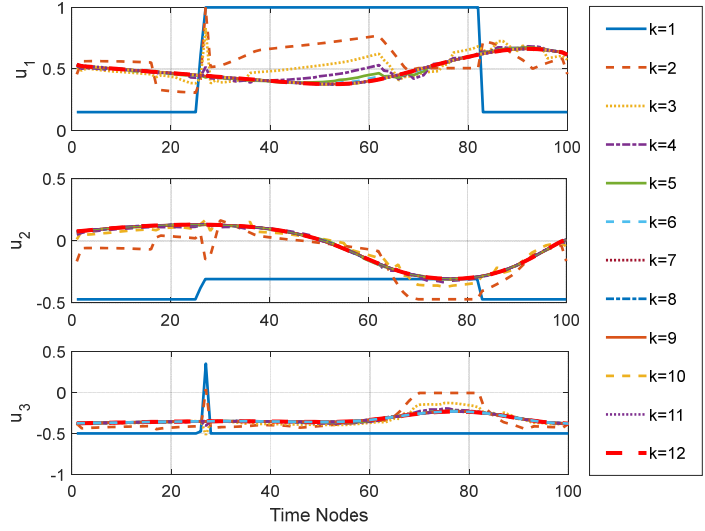

**Fig. 8 Iterative process of equivalent controls**

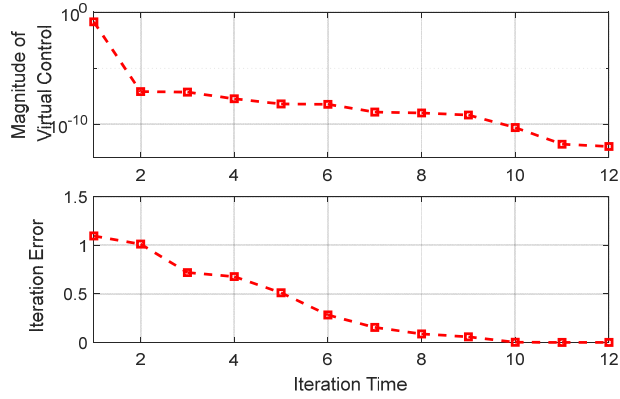

**Fig. 9 Magnitude of virtual control and iteration error in the iterative process**

In this case, we output the iterative process for the illustration of the convergence performance. The curves in **Fig. 8** illustrate the iterative process of the equivalent control defined in Eq.(8). As shown in **Fig. 8**, different color lines represent different iterative times and the red-dotted line represents the optimal convergence solution. For a sequence of arbitrarily given initial reference controls, the convergence results will finally converge to the optimal solution, which illustrates the robustness of the proposed algorithm.

The first curve of Fig. **9** illustrates the magnitude of the virtual control and the second curve expresses the variation of the iteration errors along with the iterations. In the first few iterations, the iterative process would be infeasible because of artificial infeasibility and we make a compensation with virtual control. Our greatest concern is whether the existence of virtual control will affect the authenticity of the original problem. It is obvious judging from Fig. **9** that the virtual control makes contribution only in the first two iterations and tends to zero later because of the heavy penalty in the objective function in Eq.(30). Moreover, according to the variation of the iteration errors in the second curve of Fig. **9**, convex programming with the proposed algorithm shows a good convergence performance.

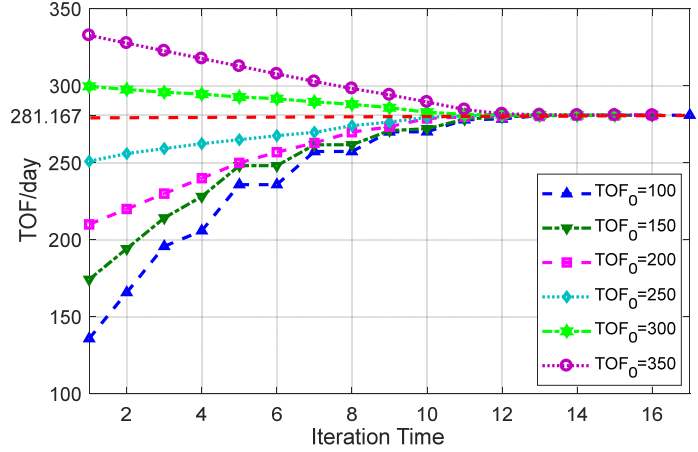

Fig. 10 Convergence performance of different initial guesses of $TOF_0$

For further discussion, different initial guesses for the TOF are specified. The lines in **Fig. 10** show the convergence performance of different initial guesses for the TOF. The guesses for TOF are specified from 100 days to 1 year and the final convergences lead to the same value of 281.167 day, which illustrates the robustness of the convex programming with the proposed algorithm.

## VI. CONCLUSION

The successive convex programming is investigated in this paper to solve the transfer trajectory optimization of solar sail. Different convexification technologies, such as change of variables, successive convexification, and virtual control, are adopted to convert the original problem into the formulation of convex programming. We present a standard successive convex approach to solve a sequence of convex sub-problems and obtain the optimal solution ultimately. Numerical test cases demonstrate the effectiveness and accuracy of the proposed algorithms. By introducing the test case from the literature, the optimality of the obtained solution is guaranteed. The results of the reintegration indicate the accuracy and validity of the solution. Compared with the traditional direct method, such as pseudospectral method, the proposed method is more efficiency. Different conditions of the initial guesses are considered in the test case. With different initial guesses for TOF, the algorithm has the same convergence solution, which illustrates the robustness of the proposed algorithm.

## VII. ACKNOWLEDGMENTS

This work is supported by the National Natural Science Foundation of China (Grant No. 11772167).

# REFERENCE

Acikmese, B., Ploen, S.R.: Convex Programming Approach to Powered Descent Guidance for Mars Landing. J. Guid. Control. Dyn. 30, 1353–1366 (2007). doi:10.2514/1.27553

Betts, J.T.: Optimal Interplanetary Orbit Transfers by Direct Transcription. J. Astronaut. Sci. 42, 247–268 (1994)

Blackmore, L., Acikmese, B., Scharf, D.P.: Minimum-Landing-Error Powered-Descent Guidance for Mars Landing Using Convex Optimization. J. Guid. Control. Dyn. 33, 1161–1171 (2010). doi:10.2514/1.47202

Bryson, A.E., Ho, Y.C.: Applied optimal control. Hemisphere Publishing, New York (1975)

Dachwald, B.: Optimization of Interplanetary Solar Sailcraft Trajectories Using Evolutionary Neurocontrol. J. Guid. Control. Dyn. 27, 66–72 (2004). doi:10.2514/1.9286

Dachwald, B., Wie, B.: Solar sail trajectory optimization for intercepting, impacting, and deflecting near-earth asteroids. AIAA Guid. Navig. Control Conf. Exhib. paper 2005-6176 (2005). doi:10.2514/6.2005-6176

Domahidi, A., Chu, E., Boyd, S.: ECOS: An SOCP solver for embedded systems. Proc. Eur. Control Conf. 3071–3076 (2013). doi:10.0/Linux-x86_64

Gong, S., Li, J.: Equilibria near asteroids for solar sails with reflection control devices. Astrophys. Space Sci. 355, 213–223 (2015)(a). doi:10.1007/s10509-014-2165-7

Gong, S., Li, J.: A new inclination cranking method for a flexible spinning solar sail. IEEE Trans. Aerosp. Electron. Syst. 51, 2680–2696 (2015)(b). doi:10.1109/TAES.2015.140117

Grant, M.C., Boyd, S.P.: The CVX Users' Guide, Release v2.1. Book. (2015). doi:10.1.1.23.5133.

Guo, J., Chu, J., Yan, J.: Establishment of UAV formation flight using control Vector parameterization and sequential convex programming. Chinese Control Conf. CCC. 2016–Augus, 2559–2564 (2016). doi:10.1109/ChiCC.2016.7553749

Harris, M.W., Açıkmeşe, B.: Minimum Time Rendezvous of Multiple Spacecraft Using Differential Drag. J. Guid. Control. Dyn. 37, 365–373 (2014). doi:10.2514/1.61505

He, J., Gong, S., Jiang, F.H., Li, J.F.: Time-optimal rendezvous transfer trajectory for restricted cone-angle range solar sails. Acta Mech. Sin. Xuebao. 30, 628–635 (2014). doi:10.1007/s10409-014-0033-x

Heiligers, J., Macdonald, M., Parker, J.S.: Extension of Earth-Moon libration point orbits with solar sail propulsion. Astrophys. Space Sci. 361, (2016). doi:10.1007/s10509-016-2783-3

Heiligers, J., Mcinnes, C.R.: Novel solar sail mission concepts for space weather forecasting. Adv. Astronaut. Sci. 152, 585–604 (2014)

Heiligers, J., Mingotti, G., McInnes, C.: Optimisation of solar sail interplanetary heteroclinic connections. In: Advances in the Astronautical Sciences. pp. 211–230 (2015)

Heiligers, J., Parker, J.S., Macdonald, M.: Novel Solar-Sail Mission Concepts for High-Latitude Earth and Lunar Observation. J. Guid. Control. Dyn. 41, 212–230 (2018). doi:10.2514/1.G002919

Hughes, G.W., Macdonald, M., McInnes, C.R., Atzei, A., Falkner, P.: Sample Return from Mercury and Other Terrestrial Planets Using Solar Sail Propulsion. J. Spacecr. Rockets. 43, 828–835 (2006). doi:10.2514/1.15889

Liu, , X., Shen, , Z., Lu, P.: Solving the Maximum-Crossrange Problem via Successive Second Order Cone Programming. Aerosp. Sci. Technol. 47, 10–20 (2015)

Liu, X.: Fuel-Optimal Rocket Landing with Aerodynamic Controls. In: AIAA Guidance, Navigation, and Control Conference (2017)

Liu, X., Lu, P.: Robust Trajectory Optimization for Highly Constrained Rendezvous and Proximity

Operations. AIAA Guid. Navig. Control Conf. (2013). doi:doi:10.2514/6.2013-4720

Liu, X., Lu, P.: Solving Nonconvex Optimal Control Problems by Convex Optimization. J. Guid. Control. Dyn. 37, 750–765 (2014). doi:10.2514/1.62110

Liu, X., Lu, P., Pan, B.: Survey on Convex Optimization in Aerospace Applications. Astrodynamics. 1, 23–40 (2017). doi:10.1007/s42064-017-0003-8

Lu, P., Liu, X.: Autonomous Trajectory Planning for Rendezvous and Proximity Operations by Conic Optimization. J. Guid. Control. Dyn. 36, 375–389 (2013). doi:10.2514/1.58436

MacNeal, R.H.: Comparison of the Solar Sail with Electric Propulsion Systems,. Comp. Sol. Sail with Electr. Propuls. Syst. (1972)

Mao, Y., Dueri, D., Szmuk, M., Açıkmeşe, B.: Successive Convexification of Non-Convex Optimal Control Problems with State Constraints. IFAC-PapersOnLine. 50, 4063–4069 (2017). doi:10.1016/j.ifacol.2017.08.789

Mao, Y., Szmuk, M., Acikmese, B.: Successive convexification of non-convex optimal control problems and its convergence properties. 2016 IEEE 55th Conf. Decis. Control. CDC 2016. 3636–3641 (2016). doi:10.1109/CDC.2016.7798816

Margraves, C.R., Paris, S.W., Hargraves, C.R.: Direct Trajectory Optimization Using Nonlinear Programming and Collocation. J. Guid. Control. Dyn. 10, 338–342 (1987). doi:10.2514/3.20223

McInnes, C.R.: Solar sail mission applications for non-keplerian orbits. Acta Astronaut. 45, 567–575 (1999). doi:10.1016/S0094-5765(99)00177-0

McInnes, C.R.: Solar Sailing: Technology, Dynamics and Mission Applications. Springer (2004)

Melton, R.: Comparison of direct optimization methods applied to solar sail problems. AIAA Pap. (2002)

Mengali, G., Quarta, A.A.: Rapid Solar Sail Rendezvous Missions to Asteroid 99942 Apophis. J. Spacecr. Rockets. 46, 134–140 (2009). doi:10.2514/1.37141

Moré, J.J., Garbow, B.S., Hillstrom, K.E.: User Guide for MINPACK-1, http://cdsweb.cern.ch/record/126569, (1980)

Nocedal, J., Wright, S.J.: Numerical optimization. Springer Series in Operations Research and Financial Engineering (2006)

Pinson, R., Lu, P.: Rapid Generation of Time - Optimal Trajectories for Asteroid Landing via Convex Optimization. Am. Astronaut. Soc. 1–19 (2015). doi:10.2514/1.G002170

Pinson, R., Lu, P.: Trajectory Design Employing Convex Optimization for Landing on Irregularly Shaped Asteroids. In: AIAA/AAS Astrodynamics Specialist Conference. pp. 1–22 (2016)

Pontryagin, L.S.: Mathematical theory in optimal control process. CRC Press (1987)

Rao, A. V.: A survey of numerical methods for optimal control. Adv. Astronaut. Sci. 135, 497–528 (2009). doi:10.1515/jnum-2014-0003

Rao, A. V., Benson, D.A., Darby, C.L., Patterson, M.A., Sanders, I., Huntington, G.T.: GPOPS: a MATLAB software for solving multiple-phase optimal control problems using the Gauss pseudospectral method. ACM Trans. Math. Softw. 37, (2010)

Song, M., He, X., He, D.: Displaced orbits for solar sail equipped with reflectance control devices in Hill's restricted three-body problem with oblateness. Astrophys. Space Sci. 361, (2016). doi:10.1007/s10509-016-2915-9

Szmuk, M., Acikmese, B., Berning, A.W.: Successive Convexification for Fuel-Optimal Powered Landing with Aerodynamic Drag and Non-Convex Constraints. AIAA Guid. Navig. Control Conf. 1–16 (2016). doi:10.2514/6.2016-0378

Tang, G., Jiang, F., Li, J.: Fuel-Optimal Low-Thrust Trajectory Optimization Using Indirect Method and Successive Convex Programming. IEEE Trans. Aerosp. Electron. Syst. 9251, 1–1 (2018). doi:10.1109/TAES.2018.2803558

Tillerson, M., Inalhan, G., How, J.P.: Co-ordination and control of distributed spacecraft systems using convex optimization techniques. Int. J. Robust Nonlinear Control. 12, 207–242 (2002). doi:10.1002/rnc.683

Vulpetti, G., Johnson, L., Matloff, G.L.: Solar Sail: A Novel Approach to Interplanetary Travel. Springer, New York (2015)

Wie, B., Murphy, D.: Solar-Sail Attitude Control Design for a Flight Validation Mission. J. Spacecr. Rockets. 44, 809–821 (2007). doi:10.2514/1.22996

Yang, H., Bai, X., Baoyin, H.: Rapid Generation of Time-Optimal Trajectories for Asteroid Landing via Convex Optimization. J. Guid. Control. Dyn. 40, 628–641 (2017). doi:10.2514/1.G002170

Zhang, S.J.: Convex Programming Approach to Real-time Trajectory Optimization for Mars Aerocapture. Aerosp. Conf. 2015 IEEE. (2015)

Zhukov, A.N., Lebedev, V.N.: Variational Problem of Transfer Between Heliocentric Orbits by Means of Solar Sail. Cosm. Res. 2, 41–44 (1964)